\documentclass{llncs}

\usepackage{amsmath}
\usepackage{epsfig}
\usepackage{subfigure}
\usepackage{times}
\usepackage[T1]{fontenc}

\newcommand{\notify}[1]{\emph{notify}(#1)}
\newcommand{\pub}[1]{\emph{pub}(#1)}
\newcommand{\sub}[1]{\emph{sub}(#1)}
\newcommand{\unsub}[1]{\emph{unsub}(#1)}
\newcommand{\true}{\text{true}}
\newcommand{\false}{\text{false}}

\begin{document}

\pagestyle{empty}

\mainmatter

\title{Preventing Coordinated Attacks Via Distributed~Alert~Exchange}

\author{
  Joaquin Garcia-Alfaro$^1$,
  Michael A.~Jaeger$^2$,
  Gero Mühl$^2$, and
  Joan Borrell$^1$
}

\institute{
  $^1$Autonomous University of Barcelona,\\
  Dept.\ of Information and Communications Engineering,\\
  Edifici Q, 08193 Bellaterra, Spain \\
  \texttt{\{jgarcia,jborrell\}@deic.uab.es}\\
  ~\\
  $^2$Berlin University of Technology,\\
  Institute for Telecommunication Systems,\\
  Communication and Operating Systems Group,\\
  EN6, Einsteinufer 17, D-10587 Berlin, Germany\\
  \texttt{\{michael.jaeger,g\_muehl\}@acm.org}
}

\maketitle

\begin{abstract}

  Attacks on information systems followed by intrusions may cause
  large revenue losses.  The prevention of both is not always possible
  by just considering information from isolated sources of the
  network.  A global view of the whole system is necessary to
  recognize and react to the different actions of such an attack.  The
  design and deployment of a decentralized system targeted at
  detecting as well as reacting to information system attacks might
  benefit from the loose coupling realized by publish/subscribe
  middleware.  In this paper, we present the advantages and
  convenience in using this communication paradigm for a general
  decentralized attack prevention framework.  Furthermore, we present
  the design and implementation of our approach based on existing
  publish/subscribe middleware and evaluate our approach for GNU/Linux
  systems. \\

  \textbf{Keywords}: Network Security, Attack Prevention System,
  Publish/Subscribe, Message Oriented Middleware, IDMEF

\end{abstract}

\section{Introduction}
\label{sec:introduction}

When attackers gain access to a corporate network by compromising
authorized users, computers, or applications, the network and its
resources can become an active part of a globally distributed or
coordinated attack.  Such an attack might be a coordinated port scan
or distributed denial of service attack against third party
networks---or even against computers on the same network.  Both,
distributed and coordinated attacks, rely on the combination of
actions performed by a malicious adversary to violate the security
policy of a target computer system.  To prevent these attacks, a
global view of the system as a whole is necessary.  Hence, different
events and specific information must be gathered and combined from
various sources to detect patterns of such a distributed attack.  This
comprises, for example, information about suspicious connections,
initiation of processes, and the creation of new files.

We already presented an attack prevention framework that is targeted
at detecting as well as reacting to distributed and coordinated attack
scenarios~\cite{garcia04icics}.  It relies on gathering and
correlating information held by multiple sources.  In this approach,
we apply a decentralized scheme based on message passing to share
alerts in a secure communication infrastructure.  This way, we can
detect and prevent those attacks performing detection and reaction
processes based on the knowledge gained through alert correlation.
In this paper, we focus on the communication infrastructure of the
attack prevention framework.  We discuss the design of its
constituting elements and the format of the exchanged messages.  We
finally evaluate a first implementation of the infrastructure deployed
on a GNU/Linux system.  The main motivation of our work aims at
fostering the collaboration between the different components of a
protection framework composed by security components in order to
achieve a more complete view of the system in whole.  Once achieved,
one can detect and react on the different actions of a coordinated or
distributed attack.

The structure of this paper is the following.  We start in
Section~\ref{sec:related-work} with a discussion of related work.
Section~\ref{sec:pub-sub-model} gives an introduction to the
publish/subscribe communication model and elaborates the convenience
in using this model for our problem domain.  In
Section~\ref{sec:related-abstraction-and-communicaction}, we briefly
overview the Intrusion Detection Message Exchange Format (IDMEF),
which is the format we use for the exchange of audit information in
our system.  We then present the cooperation scheme of the system
components in Section~\ref{sec:architecture} including a presentation
of the current state of our implementation based on xmlBlaster, an
open source publish/subscribe message oriented
middleware~\cite{ruff00xmlblaster}.
Section~\ref{sec:cidex-implementation} presents first experimental
results on the performance obtained with a first deployment of our
implementation.  We close this paper with conclusions and a discussion
of future work in Section~\ref{sec:conclusions}.

\section{Related Work}
\label{sec:related-work}

Traditional client/server solutions for security monitoring and
protection of large-scale networks rely on the deployment of multiple
sensors.  These sensors locally collect audit data and forward it to a
central server, where it is further analyzed.  Early intrusion
detection systems such as DIDS~\cite{snapp91dids} and
STAT~\cite{ilgun-et95} use this architecture and process the
monitoring data in one central node.  DIDS (Distributed Intrusion
Detection System), for instance, is one of the first systems referred
to in the literature that is using monitoring
architecture~\cite{snapp91dids}.  The main components of DIDS are a
central analyzer component called DIDS director, a set of host-based
sensors installed on each monitored host within the protected network,
and a set of network-based sensors installed on each broadcasting
segment of the target system.  The communication channels between the
central analyzer and the distributed sensors are bidirectional.  This
way, the sensors can push their reports asynchronously to the central
analyzer while the director is still able to actively request more
details from the sensors.

The issue of sensor distribution is the focus of
NetSTAT~\cite{vigna99netstat}, an application of STAT (State
Transition Analysis Technique)~\cite{ilgun-et95} to network-based
detection.  It is based on NSTAT~\cite{kemmerer97} and comprises
several extensions.
Based on the attack scenarios and the network fact modeled as a
hyper-graph, NetSTAT automatically chooses places to probe network
activities and applies an analysis of state transitions.  This way, it
is able to decide what information is needed to collect within the
protected network.  Although NetSTAT collects network events in a
distributed way, it analyzes them in a centralized fashion similarly
to DIDS.

The main limitation of both~DIDS and~NetSTAT is that their exchange of
audit data can quickly become a bottleneck due to saturation problems
associated with their centralized analyzers.  Their monitoring schemes
are straightforward as they simply push the data to a central node and
perform the computation there.  Both approaches try to reduce the
audit data sent over the network to the central analysis unit by
filtering removing information of no interest from the audit stream
and applying compression schemes afterwards.  Unfortunately, an
efficient data reduction scheme capable of forwarding only relevant
data for arbitrary threat scenarios seems infeasible.  Thus, those
systems are not able to avoid unnecessary overhead which may lead to
an overload on the central analyzer in case too many sensors are
deployed.  Furthermore, having only one single analyzer also induces
issues with respect to availability: If the central analyzer crashes
or becomes the victim of a denial of service (DoS) attack, the whole
system is completely blinded.

Some approaches published later try to solve those disadvantages.
GrIDS~\cite{stanifordchen96grids}, EMERALD~\cite{porras97emerald}, and
AAfID~\cite{spafford-zamboni2000}, for example, propose the use of
layered structures, where data is locally pre-processed and filtered,
and further analyzed by intermediate components in a hierarchical
fashion.  The computational and network load is distributed over
multiple analyzers and managers as well as over different domains to
analyze.  The analyzers and managers of each domain perform their
detection for just a small part of the whole network.  They forward
the processed information to the entity that is on the top of the
hierarchy,i.e., a master node which finally analyzes all the reported
incidents of the system.

On the one hand, GrIDS (Graph-based Intrusion Detection System for
large networks) is an evolution of DIDS~\cite{snapp91dids} and aims at
large-scale distributed systems.  It performs detection of distributed
scans and worms by aggregating computer and network information into
activity graphs~\cite{stanifordchen96grids}.  In contrast to the
centralized approach of DIDS, GrIDS allows the construction of
activity graphs that only represent hosts and the network activity
between them.  Each node of the graph represents a single host or a
group of nodes, and the edges represent network traffic between nodes.
The audit data of GrIDS is collected by means of both host- and
network-based sensors, and then forwarded to the graph manager, which
further feeds the collected information into the graph.  The whole
system deploys several graphs and graph managers in a hierarchical
fashion in order to increase the scalability of the whole system.
Therefore, each manager controls just a subset of the whole graph.
Unfortunately, only little details are provided regarding the
communication infrastructure for the exchange of information between
components which makes it hard to further analyze this system.

Similar to GrIDS, EMERALD (Event Monitoring Enabling Responses to
Anomalous Live Disturbances) extends the work of IDES (Intrusion
Detection Expert System)~\cite{lunt90ides} and NIDES (Next-Generation
Intrusion Detection Expert System)~\cite{anderson95nides} by
implementing a recursive framework in which generic building blocks
can be deployed in a hierarchical fashion~\cite{porras97emerald}. It
combines host- and network-based sensors as well as anomaly- and
misuse-based analyzers.  EMERALD focuses on the protection of
large-scale enterprise networks that are divided into independent
domains, each one of them with its own security policy.  The authors
claim to rely on a very efficient communication infrastructure for the
exchange of information between the system components.  Unfortunately,
they also provide only few details regarding their implementation.
Thus, a general statement regarding the performance of their
infrastructure cannot be made.

The AAfID (Architecture for Intrusion Detection using Autonomous
Agents) also pre\-sents a hierarchical approach to remove the
limitations of centralized approaches and, particularly, to provide
better resistance to denial of service attacks
~\cite{spafford-zamboni2000}.  It consists of four main components
called agents, filters, transceivers, and monitors organized in a tree
structure, where child and parent components communicate with each
other.  The communication subsystem of AAfID exhibits a very
simplistic design and does not seem to be resistant to a denial of
service attack as intended.  Although the set of agents may
communicate with each other to agree upon a common suspicion level
regarding every host, all relevant data is simply forwarded to
monitors via transceivers and demands for human interaction in order
to detect distributed intrusions.   

Although hierarchical approaches mitigate some weaknesses inherent to
centralized schemes, they still do not avoid bottlenecks, scalability
problems, and fault tolerance issues due to vulnerabilities at the
root level.
The first reason for this lies in the massive amount of audit data
forwarded to the higher level components which cannot be reduced
significantly through pre-filtering within small network domains.
The second reason is the centralized root domain component which may
crash or become unavailable, rendering the whole system unusable this
way.
In order to solve these problems with both central and hierarchical
data analysis, a decentralized scheme free of dedicated processing
nodes is needed.

Some decentralized message passing designs try to remove the
limitations and disadvantages of centralized and hierarchical
approaches identified above.  Their approach of distributing the
detection process has some advantages compared to centralized and
hierarchical approaches.  Mainly, decentralized architectures have no
single point of failure and bottlenecks can thus be avoided.  Some
message passing designs such as CSM~\cite{white96CSM} and
Quicksand~\cite{kruegel02distributed} try to eliminate the need for
dedicated elements by introducing a peer-to-peer architecture.
Instead of having a central monitoring station to which all data has
to be forwarded, there are independent uniform working entities at
each host performing similar basic operations.  To detect coordinated
and distributed attacks, the different entities collaborate on the
detection activities and cooperate to perform a decentralized
correlation algorithm.

These designs seem to be a promising technology to implement
decentralized architectures for the detection of attacks.  However,
the presented systems still exhibit very simplistic designs and suffer
from several practical limitations.  For instance, in some of them,
every node has to have complete knowledge of the system: All nodes
have to be connected to each other which can make the matrix of the
connections that are used for providing the alert exchanging service
grow explosively and become very costly to control and maintain.
Another important disadvantage present in this design is that the
different entities always need to know where a received notification
has to be forwarded to (similar to a queue manager).  This way, when
the number of possible destinations grows, the network view can become
extremely complex limiting the scalability of this approach.  Other
designs are based on flooding which furthermore makes the system
easier to maintain on the cost of scalability as the message
complexity quickly grows with the number of nodes in the system.

Most of these limitations can be solved efficiently by using a
distributed publish/sub\-scribe middleware.  The advantages of
publish/subscribe communication for our problem domain over other
communication paradigms is that it keeps the producers of messages
decoupled from the consumers and that the communication is
information-driven.  This way, it is possible to avoid problems
regarding the scalability and the management inherent to other designs
by means of a network of publishers, brokers, and subscribers.  A
publisher in a publish/subscribe system does not need to have any
knowledge about any of the entities that consume the published
information since the communication is anonymous. Likewise, the
subscribers do not need to know anything about the publishers.  New
services can simply be added without any impact on or interruption of
the service to other users.
In~\cite{garcia05intellcom,garcia05carnahan}, we presented an
infrastructure inspired by the decentralized architectures discussed
with the focus on removing the discussed limitations.  In the
following sections, we present further details on our work.

\section{Publish/Subscribe Model}
\label{sec:pub-sub-model}

The publish/subscribe communication model implies many-to-many
communication and is often implemented asynchronously.  It is well
suited for distributed systems~\cite{efgk03} and often used in
situations where a message (often referred to as a \emph{notification}
in the literature) published by a single entity is sent to multiple
receivers that expressed their interests previously.
Publish/subscribe systems allow for efficient and comfortable
information dissemination to receivers that may have individual
interests in arbitrary subsets of the messages published.  In contrast
to multicast communication, clients have the possibility to describe
the events they are interested in more flexible than with subscribing
to a multicast group (e.g., based on the contents of the
notification).  Clients acting as \emph{subscribers} can choose to
subscribe and later unsubscribe to filters matching a set of messages
as time goes by, while all subscribers are independent of each other.
Clients that publish notifications are called \emph{publishers}.

\subsection{Publish/Subscribe Systems}
\label{sec:pub-sub-systems}

A publish/subscribe system that implements the publish/subscribe model
consists of clients and a \emph{notification service}, the clients are
connected to.  The latter is responsible of forwarding notifications
from publishers to all interested subscribers and consists of at least
one \emph{broker} in a centralized implementation.  For scalability
reasons, it is common to implement the notification service in a
distributed fashion with a \emph{broker overlay network} that consists
of multiple brokers that cooperate to provide the notification
service.

The notification service provides a distributed infrastructure for
notification routing which includes the management of subscriptions
and the dissemination of notifications in a possibly asynchronous way.
Clients can publish notifications and subscribe to filters that are
matched against the notifications forwarded through the broker
network.  If a broker receives a new notification it checks if there
is a local client that has subscribed to a filter that matches this
notification.  If so, the message is delivered to this client.
Additionally, the broker forwards the message to neighbor brokers
according to the applied routing algorithm. We refer
to~\cite{muhl02phdthesis} for more details on publish/subscribe
systems.

\begin{figure*}[htpb]
 \begin{center}
    \subfigure[Simple publish/subscribe system.\label{fig:simple-pub-sub}]{
      \epsfig{file=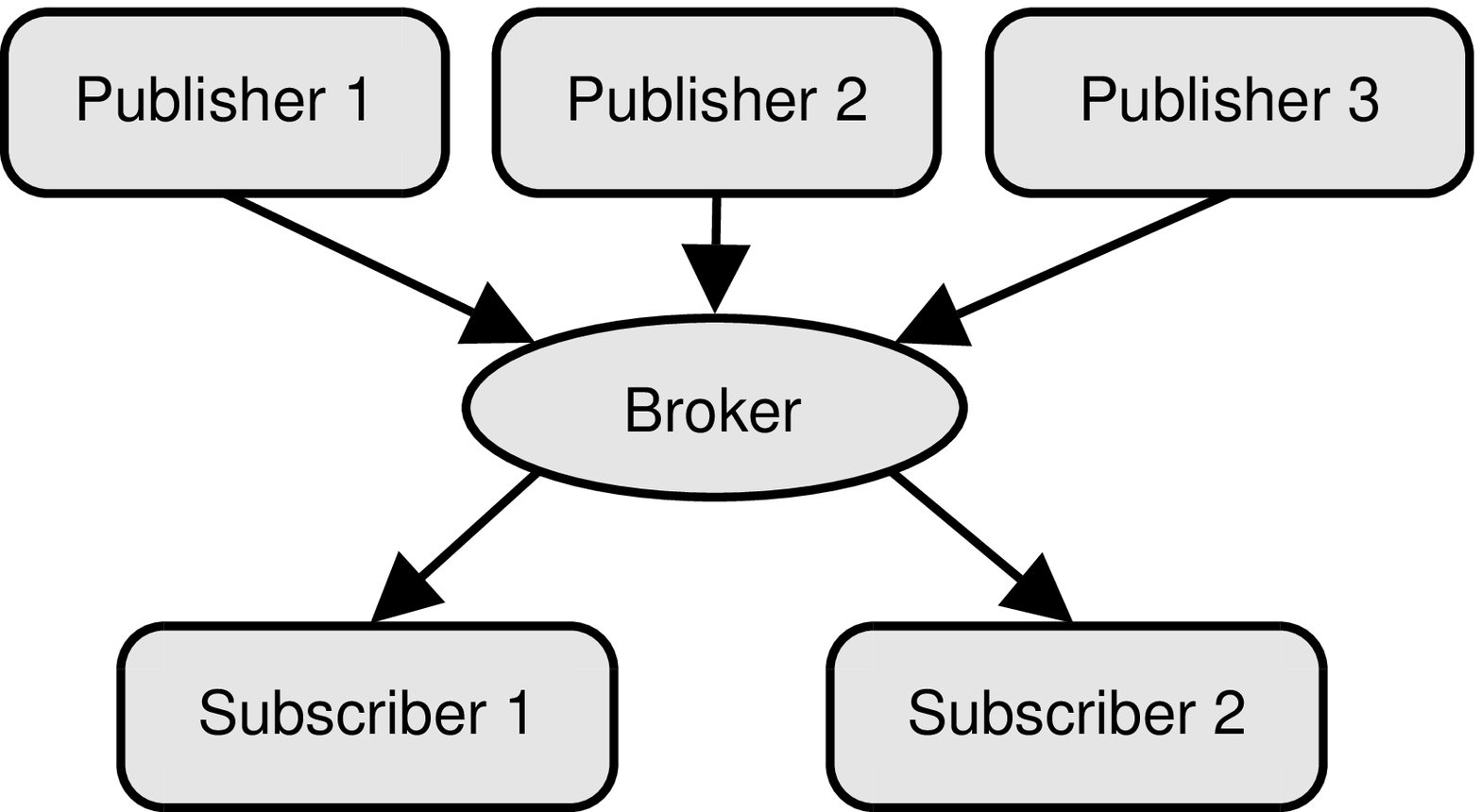,width=0.45\textwidth}
    }
    \subfigure[Extended pub/sub system.\label{fig:extended-pub-sub}]{
      \epsfig{file=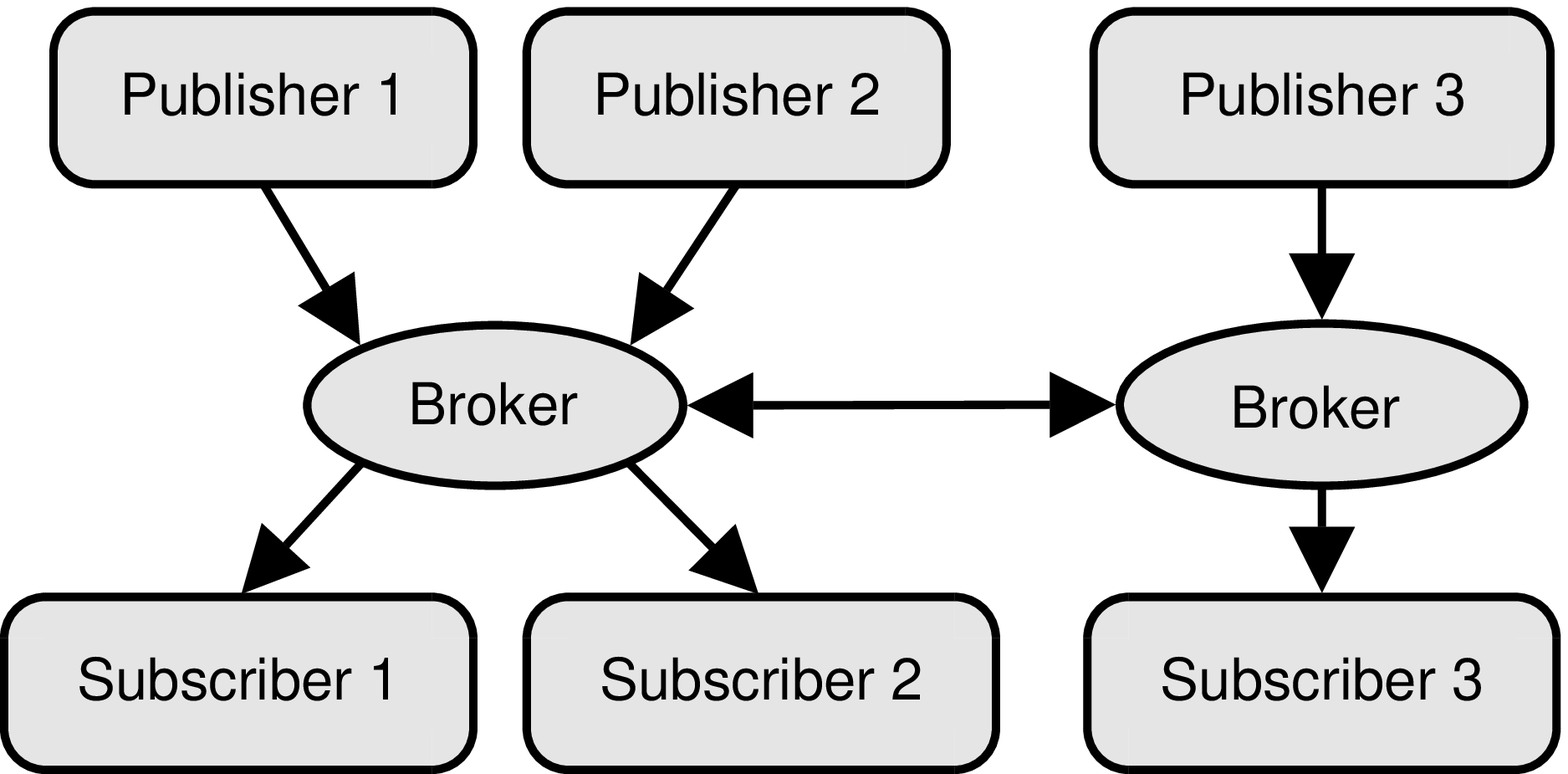,width=0.5\textwidth}
    }
    \caption{Examples of simple publish/subscribe topologies.}
    \label{fig:pub-sub-models}
 \end{center}
\end{figure*}

An example of a basic centralized publish/subscribe system is shown in
Figure~\ref{fig:simple-pub-sub}.  Here, five clients are connected to
a single broker: three clients that are publishing notifications and
two clients that are subscribed to a subset of the notifications
published on the broker.  Subscribers can choose to subscribe to the
notifications available through the broker or cancel existing
subscriptions as needed.  The broker matches the notifications it
received from the publishers to the subscriptions, ensuring this way
that every publication is delivered to all interested subscribers.

\begin{figure}[htbp]
  \centering
  \epsfig{file=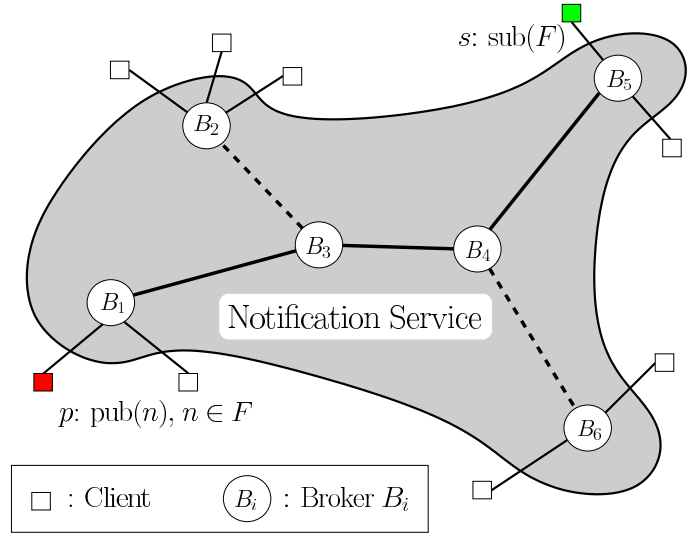,width=9cm}
  \caption{Example of a distributed publish/subscribe topology.}
  \label{fig:complete-pubsub}
\end{figure}

This very basic publish/subscribe setup can be extended by connecting
multiple brokers (cf.~Figure~\ref{fig:extended-pub-sub}), enabling
them to exchange messages.  The extended design allows subscribers on
one of the brokers to receive messages that have been published on
another broker, further freeing the subscriber from the constraints of
connecting to the same broker the publisher is connected to.  Most
available implementations make this transparent for the programmer by
keeping the same interface operations as in the centralized design.
This way, an application can easily be distributed.  In
Figure~\ref{fig:complete-pubsub}, for instance, we show a distributed
publish/subscribe topology, where a client~$p$ publishes a
notification~$n$ that is matched by filter~$F$ client~$s$ subscribed
to.  The notification service then takes care of forwarding the
notification properly over the links drawn solidly.

Regarding the subscriptions, clients are able to formulate their
interests based on the contents of the notifications or a special
attribute they carry. 
\emph{Topic-based publish/subs\-cribe} systems represent the first
variant of the publish/subscribe communication model.  Here,
publishers publish messages with respect to a topic and subscribers
specify their interest in a topic and receive all messages published
on this topic.  Topic-based subscriptions are, in turn, easier to
handle than content-based subscriptions.  Since topics can be seen as
groups in group communication~\cite{powell1996gc}, topic-based
subscription may efficiently be built on top of a group communication
mechanism such as, for example, IP~multicast~\cite{deering1989hei}.
Thus, topics are equivalent to \emph{channels}.
An extension of the topic-based approach is \emph{subject-based
publish/subscribe}.  Taking this approach, it is possible to arrange
topics in a hierarchy (subject tree) such that subscriptions not only
match notifications if the topics are the same, but also if the topic
of the subscription is an ancestor of the notification topic in the
subject tree.  In this case, a subject becomes equivalent to a
\emph{theme}.  In \emph{type-based} publish/subscribe, notifications
are equivalent to objects which allows for an easier integration in to
object-oriented programming languages.  Furthermore, it is possible
with this approach to support multiple inheritance (depending on the
programming language).

\emph{Content-based publish/subscribe} systems allow filers to work on
the content of notifications.  This way, in content-based selection
the structure of a subscription is not restricted to a topic or a
subject---it can be any function over the content of a notification.
A subscription can, thus, be formulated extremely fine-grained based
on the content of notifications using a query language that can be
arbitrarily complex.  Moreover, there does not need to be a
system-wide agreement on the set of topics as it is practically
required for topic based routing.  

Content-based subscriptions usually depend on the structure of the
message.  This can be binary data, name/value pairs, semi-structured
data, or even programming language classes containing executable code.
A subscription is often expressed in a subscription language that
specifies a filter expression over messages.

For our work, we use content-based subscription over messages with
semi-structured data.  We propose the use of XML for the structure of
a message as well as the application of XPath as the subscription
language to specify filter expressions
(cf.~Section~\ref{sec:architecture}).  In the following, we give an
outlook on the main properties of the format built on top of the XML
structure of our messages.

\section{Representation of Messages}
\label{sec:related-abstraction-and-communicaction}

In order to exchange audit information in a standard manner, two main
specifications have been considered in our job. The Common Intrusion
Specification Language (CISL), on the one hand, which was initially
proposed to allow the components of the Common Intrusion Detection
Framework (CIDF) to exchange data in semantically well-defined
ways~\cite{feirtag99cisl}. The Intrusion Detection Message Exchange
Format (IDMEF), on the other hand, was proposed by the IETF's
Intrusion Detection Exchange Format Working Group (IDWG) to accomplish
similar purposes~\cite{debar07idmef}.

Our approach is based on the IDMEF format for three main reasons.
First, this format is the basis for the similarity operator used on
the aggregation and fusion phases of our alert correlation approach
presented in~\cite{garcia04icics}. Second, there is a significant
number of tools and implementations based on the IDMEF format, such
as~\cite{libidmef}, which reduces the efforts of integrating it into
our work.  Third, the exchange of messages between the components of
our framework is compliant with the intrusion detection framework
proposed by the IDWG. Besides that, IDMEF allows the specification of
messages generated by different network security components, such as
\textit{firewalls} and \textit{network intrusion detection systems}
(NIDSs), and it can be extended to incorporate additional data
information, such as diagnoses and counter-measures, inside their
proposed format.

\begin{figure}[h]
\begin{center}
  \epsfig{file=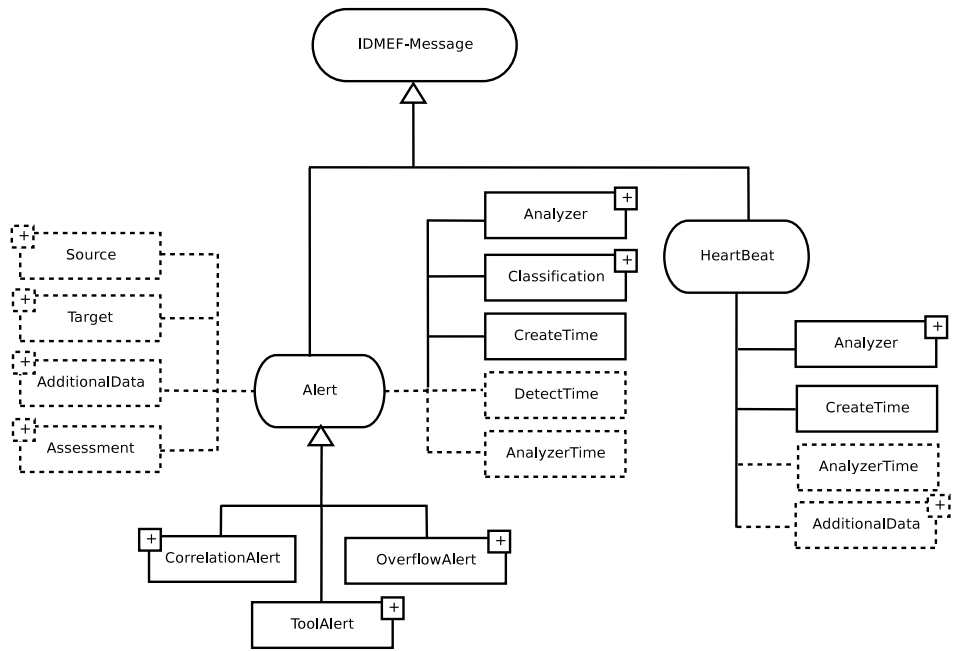,width=12.5cm}
\end{center}
\caption{The IDMEF's message class.\label{IDMEF_message_class}}
\end{figure}

Up to now, IDMEF is an \textit{internet draft} approved by the IESG
(Internet Steering Group) as an IETF RFC (Request For Comments). It is
represented in an object-oriented fashion. The class hierarchy of
IDMEF has been represented by using the Extensible Markup Language
(XML). The rationale for choosing XML is explained
in~\cite{debar07idmef}, as well as some examples of using IDMEF to
describe IDS's alerts and the IDMEF's associate Document Type
Definition (DTD)---although one may still find the current version of
IDMEF defined by using DTDs, the authors also offer a new definition
that uses XML Schemas instead of DTDs.

In Figure~\ref{IDMEF_message_class}, we show the two main types of
messages supported by IDMEF: \textit{heartbeats} and \textit{alerts}.
Heartbeats, on the one hand, are periodic messages between components,
in order to inform each other that they are operational. Alerts, on
the other hand, carry audit information, such as the component that
produced it, the classification of the detected activity, the source
and target ports related to this activity, and other optional data. In
the following, we discuss the main properties of IDMEF's alert class,
regarding aspects relevant to our work such as determining the
component which created the message, the time in which the message was
created, and the kind of activity the message is pointing out.

We start by giving an overview of the analyzer class which identifies
the component from which the message originates. Only one component is
encoded for each message, i.e., the one from which the message
originated. The class is composed, in turn, of three aggregate
classes: \textit{node}, which includes information about the node on
which the component resides; \textit{process}, which holds information
about the process in which the component is executing; and
\textit{analyzer}, which carries information about other components
which, in turn, forwarded the original information.

The rationale behind the recursive aggregation of component's
references within the IDMEF's analyzer class is that when a component
receives an IDMEF alert and wants to forward it to another component,
it needs to substitute the original component information with its own
since, as we pointed out above, just one component is encoded for each
message. This way, and in order to preserve the original component
information, it may be included in the new component definition as a
reference to the previous component. This mechanism will allow
component path tracking.

The class analyzer has eight attributes: \textit{analyzerid},
\textit{name}, \textit{manufacturer}, \textit{model},
\textit{version}, \textit{class}, \textit{ostype}, and
\textit{osversion}. The \textit{manufacturer}, \textit{model},
\textit{version}, and \textit{class} attributes' contents are
vendor-specific, but may be used together to identify different types
of components. The \textit{ostype} and \textit{osversion} attributes'
contents are, respectively, the operating system name and the
operating system version in which the component's process is executed.
Finally, the \textit{analyzerid} and \textit{name} attributes'
contents provide, respectively, the unique identifier and the explicit
name for the component in the system.

Regarding the timestamps of a message, the IDMEF standard defines the
following three different classes to represent time: (1)
\textit{CreateTime}, which is the time when the message is created by
a component; (2) \textit{DetectTime}, which is the time when the event
or events that caused the creation of a message were detected; (3)
\textit{AnalyzerTime}, which is the time when the original component
forwarded this message. The final object for each instance contains
information such as the number of seconds since the \textit{epoch},
the local GMT offset, and the number of microseconds.  Even though all
the three timestamps can be provided by each component when generating
a message, just the one defined by the \textit{CreateTime} class is
considered mandatory by the IDMEF standard.

The classes source and target contain, respectively, information about
the possible origin and destination of the events that motivated the
generation of the message. An event may have more than one source
(e.g., a distributed denial of service attack), more than one target
(e.g., a port sweep). Both, source and target classes, are composed of
information about the \textit{node}, the \textit{user}, the
\textit{process}, and the \textit{network service} that motivated the
message. The target class includes, moreover, a list of affected
\textit{files}. Referring to their attributes, both source and target
classes have the following two common attributes: (1) \textit{ident},
which is a unique identifier for either the source or target class;
(2) \textit{interface}, which may be used by a component multiple
interfaces to indicate which interface this source or target was seen
on. Furthermore, the class source includes the attribute
\textit{spoofed}, which indicates whether the source is, as far as the
component can determine, a spoofed address. Similarly, the class
target includes the attribute \textit{decoy}, to indicate whether the
target is, as far as the analyzer can determine, a decoy.

The classification class contains the \textit{name} of the event that
motivated the creation of a message, or other information which allows
the components to determine what the message is pointing out. It is
composed of one aggregate class, the class \textit{reference}, which
contains information about external documentation sites, that will
provide background information about such an event. Similarly, the
assessment class is used to provide the component's assessment of an
event, and it is composed of information about the \textit{impact},
\textit{actions} that may be taken in response, and a measurement of
the \textit{confidence} the component has in its evaluation of the
event.

Finally, the IDMEF's alert class can be augmented with additional
information by means of the aggregate classes \textit{AdditionalData},
\textit{CorrelationAlert}, \textit{ToolAlert}, and
\textit{OverflowAlert}. The information aggregated by those classes is
often useful in order to associate different messages pointing out to
similar activities---and reported by different components---as well as
to extend the standard IDMEF model with additional features, such as
complex data types and relationships. The \textit{AdditionalData}
class, first, includes information that does not fit into
the IDMEF's data model. This may be an atomic piece of data, or a
large amount of data. The \textit{CorrelationAlert} class, on the
second hand, may include additional information related to the
correlation process in which this message is involved. The
\textit{OverflowAlert} and \textit{ToolAlert} classes, on the third
hand, include, respectively, information related to buffer overflow
attacks, and information related to the use of attack tools or other
malevolent programs (e.g., \textit{trojan horses}, \textit{rootkits},
and so on).

\section{Communication Infrastructure}
\label{sec:architecture}

In this section we give an outlook to the operational details of the
communication infrastructure presented
in~\cite{garcia04icics,garcia05intellcom}. As our motivation is not
targeted at developing a new publish/subscribe system, we try to reuse
as much available code and tools as possible. For our experiments
(cf.~Section~\ref{sec:cidex-implementation}) we used
\emph{xmlBlaster}, an open source publish/subscribe message oriented
middleware~\cite{ruff00xmlblaster}. It connects a set of nodes that
build up the infrastructure for exchanging alerts using the interface
operations offered by the underlying middleware. Each xmlBlaster
message consists of a header filtering that can be applied to, a body,
and a system control section. The body of an xmlBlaster message is
formulated using IDMEF format
(cf.~Section~\ref{sec:related-abstraction-and-communicaction}).
Filters are XPath expressions that are evaluated over the message
header to decide if a message has to be delivered to a subscriber. We
discuss the essential interface operations offered by xmlBlaster in
the following section.

\subsection{Interface Operations}
\label{sec:interface}

\begin{figure}[htbp]
  \centering
  \epsfig{file=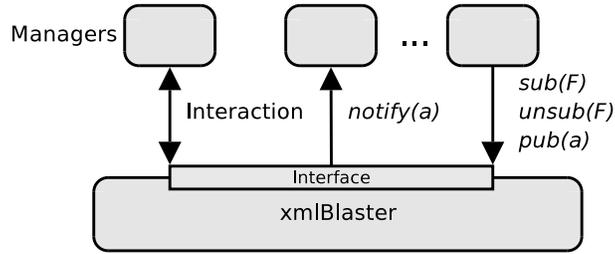,width=8cm}
  \caption{Black box view of a publish/subscribe system.}
  \label{fig:blackboxpubsubsystem}
\end{figure}

Conceptually, the alert communication infrastructure offered through
xmlBlaster can be viewed as a black box with an \emph{interface}
(cf. Figure~\ref{fig:blackboxpubsubsystem}). It offers a number of
\emph{operations}, each of which may take a number of
\emph{parameters}. Clients can invoke \emph{input operations} from the
outside, and the system itself invokes \emph{output operations} to
deliver information to clients. To publish alerts, clients invoke the
\pub{$a$} operation, giving the alert $a$ as parameter. The published
alert can potentially be delivered to all clients connected to the
system via an output operation called \notify{$a$}. Clients register
their interest in specific kinds of alerts by issuing subscriptions
via the \sub{$F$} operation, which takes a filter $F$ as parameter.
Each client can have multiple active subscriptions which must be
revoked separately by using the \unsub{$F$} operation.

All these operations are instantaneous and take parameters from the
set of all clients $\mathcal{C}$, set of all alerts $\mathcal{A}$, and
the set of all filters $\mathcal{F}$. Formally, a filter $F \in
\mathcal{F}$ is a mapping defined by
\begin{equation}
  F: \quad a\,\longrightarrow\,\{\true,\false\} \qquad \forall a \in
  \mathcal{A}
  \label{eq:filter-mapping}
\end{equation}
We say that a \emph{notification $n$ matches filter $F \in \mathcal{F}$}
iff $F(a)=\true$.  We require that each alert can only be
published once and that every filter is associated with a unique
identifier in order to enable the alert communication infrastructure
to identify a specific subscription.

\subsection{Components and Interactions}

\begin{figure}[hbtp]
  \centering \epsfig{file=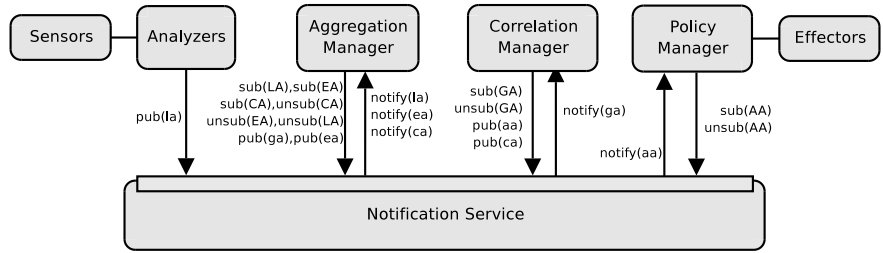,width=.95\textwidth}
  \caption{Overview of the main components and their interactions.}
 \label{fig:preventioncellsframework}
\end{figure}

As shown in Figure~\ref{fig:preventioncellsframework}, and according
to the general framework introduced in~\cite{garcia04icics}, each node
of the architecture is made up of a set of \emph{local analyzers}
(with their respective detection units or sensors), a set of
\emph{alert managers} (to perform alert processing and manipulation
functions), and a set of \emph{local reaction units} (or effectors).
These components, the interactions between them, and the alert
communication infrastructure, are described in the following.

\emph{Analyzers} are local elements which are responsible for
processing local audit data. They process the information gathered by
associated sensors to infer possible alerts. Their task is to identify
occurrences which are relevant for the execution of the different
steps of an attack and pass this information to the correlation
manager via the publish/subscribe system. They are interested in
\emph{local alerts} which are detected in a sensor's input stream and
published through the publish/subscribe system by invoking the
\pub{$la$} operation, giving the local alert $la$ as parameter.  Local
alerts are exchanged using IDMEF messages
(cf.~Section~\ref{sec:related-abstraction-and-communicaction}).

Each notification~$la$ has a unique classification and a list of
attributes with their respective types to identify the analyzer that
originated the alert (\emph{AnalyzerID}), the time the alert was
created (\emph{CreateTime}), the time the event(s) leading up to the
alert was detected in the sensor's input stream (\emph{DetectTime}),
the current time on the analyzer (\emph{AnalyzerTime}), and the
source(s) and target(s) of the event(s) (\emph{Source} and
\emph{Target}).  All possible classifications and their respective
attributes must be known by all system components, i.e., sensors,
analyzers, and managers, and all analyzers are capable of publishing
instances of local alerts of arbitrary types.

\emph{Managers} are the components in charge of performing aggregation
and correlation of local alerts and external events. As pointed out
in~\cite{garcia04icics}, the use of multiple analyzers and sensors
together with heterogeneous detection techniques increases the
detection rate, but it also increases the number of information to
process. In order to reduce the number of false negatives and
distribute the load that is imposed by the alerts, our architecture
provides a set of aggregation and correlation \emph{managers}, which
perform aggregation and correlation of both, local alerts (i.e.,
messages provided by the node's analyzers) and external messages
(i.e., the information received from other collaborating nodes). In
the following, we describe the basic interactions of the two main
managers: \textit{aggregation} and \textit{correlation} managers.

\paragraph{Aggregation Manager.}

The basic functionality of each aggregation manager is to cluster
alerts that correspond to the same occurrence of an
action~\cite{garcia04icics}. Each aggregation manager registers its
interest in a subset~$\mathcal{L}_A$ of local alerts published by
analyzers on the same node by invoking the \sub{$LA$} operation, which
takes the filter~$LA$ as parameter, with
\begin{equation}
  LA(a) = \left\{
    \begin{array}{lll}
      \true & , & a \in \mathcal{L}_A \\
      \false & , & \text{otherwise.}
    \end{array}
    \right.
  \label{eq:la-def}
\end{equation}

Similarly, the aggregation manager also registers its interest in a
set of related external alerts~$\mathcal{E}_A$ by invoking the
\sub{$EA$} operation with filter~$EA$ as parameter, and
\begin{equation}
  EA(a) = \left\{
  \begin{array}{lll}
    \true & , & a \in \mathcal{E}_a \\
    \false & , & \text{otherwise.}
  \end{array}
  \right.
\end{equation}

Finally, it registers its interest in local correlated
alerts~$\mathcal{C}_A$ by invoking the \sub{$CA$} operation with
\begin{equation}
  CA(a) = \left\{
  \begin{array}{lll}
    \true & , & a \in \mathcal{C}_A \\
    \false & , & \text{otherwise.}
  \end{array}
  \right.
\end{equation}

Once subscribed to these three filters, the communication
infrastructure will notify the subscribed managers of all matching
alerts via the output operations \notify{$la$}, \notify{$ea$} and
\notify{$ca$} with $la \in \mathcal{L}_A$, $ea \in \mathcal{E}_A$ and
$ca \in \mathcal{C}_A$. All notified alerts are processed and,
depending on the clustering and synchronization mechanism, the
aggregation manager can publish global and external alerts by invoking
\pub{$ga$} and \pub{$ea$}. Finally, it can revoke active subscriptions
separately by using the operations \unsub{$CA$}, \unsub{$EA$} and
\unsub{$LA$}.

\paragraph{Correlation Manager.}

The main task of this manager is the correlation of alerts described
in~\cite{garcia04icics,annales06}. It operates on the set of global
alerts~$\mathcal{G}_A$ published by the aggregation manager. To
register its interest in these alerts, it invokes \sub{$GA$}, which
takes the filter~$GA$ as parameter with
\begin{equation}
  GA(a) = \left\{
  \begin{array}{lll}
    \true & , & a \in \mathcal{G}_A \\
    \false & , & \text{otherwise.}
  \end{array}
  \right.
  \label{eq:ga-def}
\end{equation}

The notification service will then notify the correlation manager of
all matched alerts with the output operation \notify{$ga$}, $ga \in
\mathcal{G}_A$. Each time a new alert is received, the correlation
mechanism finds a set of action models that can be correlated in order
to form a scenario leading to an objective. It then includes this
information into the \emph{CorrelationAlert} field of a new IDMEF
message and publishes the correlated alert by invoking \pub{$ca$},
giving the notification $ca \in \mathcal{C}_A$ as parameter. To revoke
the subscription, it uses \unsub{$GA$}.

The correlation manager is also responsible for reacting on detected
security violations. The algorithm used is based on the
anti-correlation of actions to select appropriate counter-measures in
order to reconfigure, for instance, the security
policy~\cite{debar07policy}. As soon as a
scenario is identified, the correlation mechanism may look for
possible action models that can be anti-correlated with the individual
actions of the supposed scenario, or even with the goal objective.

The set of anti-correlated actions represents the set of
counter-measures available for the observed scenario. The definition
of each anti-correlated action contains a description of the
counter-measures which should be invoked (e.g., hardening the security
policy). Such counter-measures are included into the \emph{Assessment}
field of a new IDMEF message and published by invoking \pub{$aa$},
using the \emph{assessment alert} $aa$ as parameter.

Finally, a \emph{policy manager} will register and revoke its interest
in these assessment alerts by invoking \sub{$AA$} and \unsub{$AA$}.
Once notified, the policy manager may perform the post-processing of
the received alerts before sending them, for example, to a set of
associated policy reconfiguration effectors.

\section{Deployment and Evaluation}
\label{sec:cidex-implementation}

In order to evaluate the performance of our proposal, we deployed a
set of analyzers and managers publishing and receiving IDMEF messages
based on the \emph{DARPA Intrusion Detection Evaluation Data
  Sets}~\cite{lippmann00ideval}. This evaluation data set contains
more than 300 instances of 38 different automated attacks that were
launched against victim hosts in seven weeks of training data and two
weeks of test data.

The complete set of messages were published as local and external
alerts through the notification service of xmlBlaster, and then
processed and republished in turn to the set of subscribed managers.
The exchange of alerts proved to be satisfactory, obtaining a
throughput performance higher than 150 messages per second on an
Intel-Pentium M 1.4 GHz processor with 512 MB RAM, analyzers and
managers on the same machine running Linux 2.6.8, using Java HotSpot
Client VM 1.4.2 for the Java-based broker. Message delivery did not
become a bottleneck as all messages were processed in time and
the saturation point has never been reached.

The implementation of both, publishers and subscribers, was based on
the \emph{libidmef} C~library~\cite{libidmef} in order to build and
parse compliant IDMEF messages. In turn, \emph{libidmef} is built over
the libxml~library~\cite{libxml}. The libxml library provides two
interfaces to parse XML data: a DOM style tree interface, and a SAX
style event-based interface for our implementation. Up to now, we are
using the DOM interface due to its easiness of use. Its main drawback
is, however, that its memory usage is proportional to the size of the
XML data. For this reason, we are currently rewriting our
implementation to use the SAX-based interface. This would help us to
decrease the amount of memory that is currently necessary to maintain
the entire XML tree in memory.

\begin{figure*}[htpb]
  \begin{center}
    \epsfig{file=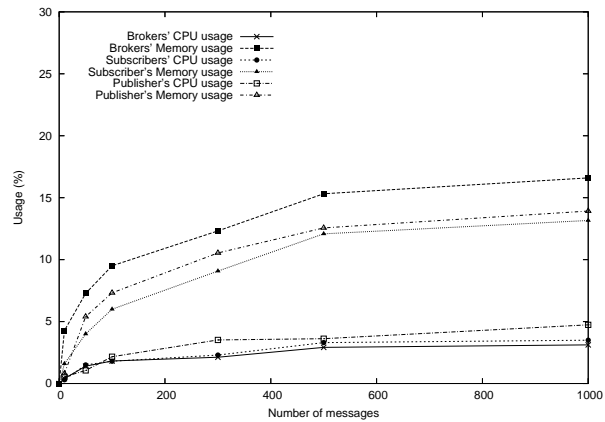,width=8cm}
    \caption{Processing and memory consumption.}
    \label{fig:broker-performance}
  \end{center}
\end{figure*}

The communication between analyzers and managers through xmlBlaster
brokers was based on the xmlBlaster internal socket protocol and
implemented using the C~socket library~\cite{ruff00xmlblaster} for
xmlBlaster, which provides asynchronous callbacks to Java-based
brokers. The managers formulated their subscriptions using XPath
expressions, filtering the messages they wished to receive from the
broker.

In Figure~\ref{fig:broker-performance}, we show the processing time
and memory space used by xmlBlaster brokers during the exchange of
alerts.  The first curve represents the percentage of CPU load used by
each broker. The second curve represents the quantity of memory used
by each broker. As we can notice in the first curve, the percentage of
processing time used by the brokers is quite stable and negligible for
the normal performance of a normal system. The second curve reflects,
however, that the cost in memory is quite high. We consider that this
consumption is due to the message management and we hope that the new
version of our prototype based on a more efficient XML parsing and
building scheme will lower it as discussed above.

\section{Conclusions}
\label{sec:conclusions}

We presented a message passing design for the exchange of audit
information between the security components of a platform for the
detection of and reaction on coordinated attacks. The design is based
on a publish/subscribe model. Instead of having a central or master
monitoring station to which all data has to be forwarded, there are
independent uniform working entities at each host performing similar
basic operations. The information gathered by each entity is
disseminated to other interested entities through a notification
service based on a publish/subscribe broker network which allows
messages to be sent via a push or pull data exchange. The main
advantage of this model for the exchange of audit information between
components is, on the one hand, that it keeps the producer of messages
separated from the consumers and, on the other hand, that the
communication is information-driven.  This way, it allows us to avoid
problems regarding the scalability and the management inherent to
other designs, by means of a network of publishers, brokers, and
subscribers. A publisher in a publish/subscribe system does not need
to have any knowledge about any of the entities that consume the
published information. Likewise, the subscribers do not need to know
anything about the publishers.  Services can be added without any
impact on or interruption of the service to other elements.

In Section~\ref{sec:related-abstraction-and-communicaction}, we
discussed the main properties of the Intrusion Detection Message
Exchange Format (IDMEF) as the format that is built on top of the XML
structure of the messages exchanged between the components of our
platform; we presented in Section~\ref{sec:architecture} the
operational details (interface operations and interaction) of our
communication infrastructure; and we discussed the initial results of
a first prototype of our approach in
Section~\ref{sec:cidex-implementation}. We think that these results
give us good hope that the use of a publish/subscribe system for the
communication infrastructure indeed increases the scalability of the
proposed architecture. Motivated bey the high memory usage, and as
pointed out in Section~\ref{sec:cidex-implementation}, we are actually
moving our current implementation to the SAX interface, since it does
not maintain the entire XML tree in memory, which means that the load
will considerably decrease.

As an extension of the work presented in this paper, we may first
consider to secure the communication partners by utilizing the SSL
protocol. This way, each node will receive a private and a public key.
The public key of each node will be signed by a certification
authority (CA) that is responsible for the protected network. Hence,
the public key of the CA has to be distributed to every node as well.
The secure SSL channel will allow the communicating peers to
communicate privately and to authenticate each other, thus preventing
malicious nodes from impersonating legal ones. The implications coming
up with this new feature, such as compromised key management or
certificate revocation, would be part of this future work. We may also
consider as further work a more in-depth study about privacy
mechanisms by exchanging alerts in a pseudonymous manner. By doing
this, one may provide the destination and origin information of alerts
(\emph{Source} and \emph{Target} field of IDMEF messages) without
violating the privacy of publishers and subscribers located on
different domains. Our study may cover the design of a pseudonymous
identification scheme, trying to find a balance between identification
and privacy. This also represents further work that remains to be
done.

\section*{Acknowledgments}

\noindent The collaboration between J. Garcia-Alfaro, F. Cuppens, and
F. Autrel sharpened many of the arguments presented in this paper. The
authors graciously acknowledge the financial support received from the
following organizations: Spanish Ministry of Science and Education,
and the Catalan Government's Agency for Management of University and
Research Grants (AGAUR).

\end{document}